\documentclass[12pt]{article}
\usepackage{amsfonts}
\usepackage{amssymb}

\begin{document}

\title{Quantization of the Smoluchowski equation
and the problem of quantum tunneling at zero temperature}

\author{ A. O. Bolivar    \vspace{3pc} \\
Instituto M\'{a}rio Sch\"{o}nberg \\
                Ceil\^{a}ndia, 72225-971, Cx. P. 7316, D.F, Brazil. }

\date{\today}

\maketitle


\begin{abstract}
In this article we address the problem of quantum tunneling of a
non-Markovian Brownian particle away from thermal equilibrium. We
calculate the Kramers escape rate at low  temperature (including
the zero temperature case) in the Smoluchowski limit (strong
friction regime). Our main findings are: (i) our quantum escape
rate is valid far from the thermal equilibrium and is
non-Markovian, but it becomes Markovian as the correlation time
vanishes; (ii) at thermodynamic equilibrium we obtain a
non-Markovian quantum  rate that predicts a superfluidity
phenomenon in the Markovian limit at low  and  zero temperatures.
\end{abstract}
\vspace{20pc}


\section{Introduction: Quantum tunneling}


Let us consider a quantum particle moving in a double-well
potential. The transition of this particle over the barrier
potential from a metastable state toward another state is known as
quantum tunneling. This phenomenon takes place in many areas of
physics (e.g., in condensed matter physics), chemistry, astronomy,
and biology \cite{Pechukas2001,Hanggi90,Chandrasekhar43}. From the
experimental standpoint
 quantum tunneling has been investigated on the basis of rate experiment
 performed in the following areas \cite{Hanggi90}: Tunneling in
 biophysical transport, quantum diffusion in solids, chemical conversion
 processes, tunneling in ferromagnetic materials, electron tunneling
 in amorphous alloys, nucleation of vortices in HeII, escape of
 electrons from the surface of liquid helium, low-temperature
 Josephson-junction systems, tunneling of protons on hydrated protein
 powders, and many others.\vspace{1pc}

In isolated systems quantum tunneling process has been studied
making use of the Schr\"odinger equation \cite{Das2001}, whereas
in open systems, i.e., systems immersed in a reservoir undergoing
Brownian motion, we cannot follow such approach since there is no
wavefunction describing a Brownian particle, for instance. The
nontrivial issue is therefore to find out a manner of elaborating
a theory of quantum transport so as to clarify the role played by
dissipation and fluctuation  on quantum tunneling. To be complete
this theory has to take into account not only equilibrium and
Markovian properties but also non-Markovian and non-equilibrium
features \cite{Pollak2005}.\vspace{1pc}

The usual Hamiltonian approach for describing open systems is as
follows. Imagine an environment consisting of a set of harmonic
oscillators coulpled to the Brownian particle. Once found the
Hamiltonian function of whole system (particle plus environment),
canonical quantization procedure could be employed. The Feynman
path integral formalism is used to derive equations of motion for
the reduced density matrix describing the quantum motion of the
Brownian particle alone. On the basis of this method Caldeira and
Leggett \cite{Caldeira83a} have quantized the Kramers equation
\cite{Kramers40} and analyzed quantum tunneling
\cite{Caldeira83b}. More recently, following the Caldeira--Leggett
approach Ankerhold  at al. \cite{Ankerhold2005} have derived a
quantum Smoluchowski equation and explored its physical meaning
applying it to chemical reactions, mesoscopic physics, and charge
transfer in molecules. Meanwhile, other works \cite{Machura2004}
have revealed some drawbacks underlying such quantum Smoluchowski
equation, since it violates  the Second Law of
Thermodynamics, for instance. From our point of view the restlessness over this
internal inconsistency leads us to search for alternative ways of
quantizing the Smoluchowski equation, thereby eschewing any  {\it ad
hoc} procedures as achieved in Ref.\cite{Machura2004}.\vspace{1pc}

Another Hamiltonian approach \cite{Banik2003} has been developed and 
used in \cite{Banerjee2003} to derive a
quantum Smoluchowski equation and investigate the issue of
quantum tunneling at zero temperature, in contrast with
Ankerhold  et al.'s survey. Such alternative way does not
depend on the path integral formalism, but is based on the
canonical quantization.\vspace{1pc}

In recent years, we have put forward a non-Hamiltonian method for
quantizing open systems \cite{Bolivar2004,Bolivar98,Bolivar2005}.
There we start directly with the stochastic dynamics (Langevin and
Fokker--Planck equations) and quantize it by making use of a
Fourier transform carrying the Planck constant. Accordingly, our
approach is independent of path integral techniques and is not
based on canonical quantization. \vspace{1pc}

In order to contribute to a general theory of quantum tunneling within a
non-Hamiltonian framework we have already started in
\cite{Bolivar2005} we organize our article as follows:

I. Introduction: Quantum tunneling

II. Our generalized Langevin equation

III. Our non-Markovian Smoluchowski equation

IV. Quantization of our Smoluchowski equation

V.  Quantum tunneling

VI. Summary and discussions

Appendix: Derivation of our Fokker--Planck equation [Eq.(13)]

\section{Our generalized Langevin equation}

As a physical model of a stochastic process we consider a particle with mass $m$
immersed into an environment.  This particle undergoing a Brownian motion 
is characterized by the  stochastic position $X=X(t)$ and the 
stochastic momentum $P=P(t)$, while the environment is specified by a random
variable $\Psi=\Psi(t)$. Such physical quantities could be intertwined 
through the relations
\begin{equation}
X=Q+\Delta Q\quad \quad;\quad \quad 
P=m\frac{d X}{d t}, 
\end{equation}
where $\Delta Q=\alpha b_1(t)\Psi(t)$, $t$ being a parameter, called time, and   
$\alpha$ a dimensional constant such
that $\Delta Q$ has dimension of length. $d/dt$  denotes a differential operator 
acting upon $X$, and  
$b_1(t)$ a time-dependent parameter measuring the strength of the 
environment effects upon the particle. We define it as being
\begin{equation}
b_1=b_1(t)=\int_{0}^{t}
\langle \Psi(t')\Psi(t'')\rangle dt'',
\end{equation}
where the mean  
$$
\langle \Psi(t')\Psi(t'')\rangle=\int \int \int \psi(t')\psi(t'') 
D_{XP\Psi}(x,p,\psi,t) dxdp d\psi=
$$
$$
\int \psi(t')\psi(t'') 
D_{\Psi}(\psi,t)d\psi
$$
 is calculated in terms of the joint probability density function 
$D_{XP\Psi}(x,p,\psi,t)$ or the probability density $D_{\Psi}(\psi,t)$. 
\vspace{1pc}

One assumes the motion
of the Brownian particle moving in an external potential $V(X)$
to be described by the stochastic differential equations 
in phase space ($X,P$), known as Langevin's equations \cite{Hanggi90,Chandrasekhar43}, 
\begin{equation}
\frac{dP}{dt}= -\frac{dV}{dX}-\frac{\gamma}{m}P+b_1\Psi
\quad;\quad \frac{dX}{dt}=\frac{P}{m},
\end{equation}
where
$-\gamma P/m$ denotes a (memoryless) frictional force
activating the particle motion.  There $\Psi$ has the statistical 
properties  
\begin{equation}
\langle \Psi(t')\Psi(t'')\rangle =2D^{1/3}\delta(t''-t')\quad \quad;
\quad \quad \langle \Psi \rangle=0,
\end{equation}
making the stochastic process Markovian. $\delta(t''-t')$ is the 
Dirac delta function and  
$D$ is a  constant -- to be determined by 
the physics of the problem -- such that $b_1\Psi=D^{1/3}$ in Eq.(3) 
has in fact dimension of newton.
\vspace{1pc}

It  is important to note that as  the environmental parameter
$b_1(t)$ does vanish, the stochastic quantities $P$
and $X$ reduce to the respective deterministic values $p=mdq/dt$
and $x=q$, provided $D_{XP}(x,p)=\delta(x-q)\delta(p-p')$. 
Physically, that means that the initially open system becomes isolated 
from its environment and turns out to be described by Newton's equations 
\begin{equation}
\frac{dp}{dt}=-\frac{dV(x)}{dx} -\gamma \frac{p}{m} \quad \quad;\quad \quad 
\frac{dx}{dt}=\frac{p}{m}.
\end{equation}
For this reason one says that the Langevin equations (3) are a generalization 
of Newton's equations (5).
\vspace{1pc}

In the literature \cite{Hanggi90}  
the non-Markovian character is
introduced by means of the following statistical properties of
$\Psi$
\begin{equation}
\langle \Psi(t')\Psi(t'')\rangle =(D/t^2_c)^{1/3}e^{-(t''-t')/t_c}
\quad \quad;\quad \quad
\langle \Psi \rangle=0,
\end{equation}
where $t''>t'$ and $t_c$ is the correlation time between the Brownian particle 
and its environment. One takes into account a memory friction kernel in the 
Langevin equations (3): 
\begin{equation}
\frac{dP}{dt}= - \frac{d V}{d X}-
\int_{0}^{t}\beta(t-\tau)\frac{P(\tau)}{m}d\tau+b_1\Psi
\quad;\quad \frac{dX}{dt}=\frac{P}{m}.
\end{equation}
Both the frictional kernel $\beta(t-\tau)$ and the 
fluctuating function $\Psi(t)$ are coupled 
by means of the dissipation-fluctuation theorem \cite{Zwanzig73}
$$
\langle \Psi(t')\Psi(t'')\rangle =\kappa_B T\beta(t-\tau).
$$ 
Physically, such a theorem assures that the Brownian particle will 
always attain the thermal equilibrium of the heat bath 
characterized by Boltzmann's constant $\kappa_B$ and the temperature 
$T$. 
As $\beta(t-\tau)=2\gamma \delta(t-\tau)$ and 
the correlation time $t_c$ tends to zero, i.e., 
$t_c \rightarrow 0$, the expression  
(6) reduces to (4) while (7) reproduces  (3). 
Thereby, the stochastic dynamics (7), along with the statistical 
properties (6), are called the generalized Langevin equations 
\cite{Zwanzig73}.
\vspace{1pc}

In the present  paper 
our purpose is to make another extension of the Langevin approach. 
To begin with, we hold the definition of $X$ in (1) and  generalize 
the stochastic momentum $P=dX/dt$ according to 
\begin{equation}
\bar P=P+\Delta P,
\end{equation}
where $\Delta P=-mb_2(t)\Psi(t)$, with $b_2(t)$ defined as
\begin{equation}
b_2=b_2(t)=\int_{0}^{t}\langle \Psi(t')\rangle dt'.
\end{equation}
Accordingly, the Langevin equations (3) turn out to be written as 
\begin{equation}
\frac{d\bar P}{dt}= -\frac{dV}{dX}-\frac{\gamma}{m}\bar P+b_1\Psi
\quad;\quad \frac{dX}{dt}=\frac{\bar P}{m}+b_2\Psi,
\end{equation}
in phase space $(X,\bar P)$, with 
\begin{equation}
\langle \Psi(t')\Psi(t'')\rangle =(D/t^2_c)^{1/3}e^{-(t''-t')/t_c}\quad \quad;
\quad \quad \langle \Psi \rangle=(C/t_c^2)^{1/3}e^{-t'/t_c}, 
\end{equation}
and 
\begin{equation}
b_1=(Dt_c)^{1/3}(1-e^{-t/t_c})\quad \quad;\quad \quad
b_2=(Ct_c)^{1/3}(1-e^{-t/t_c}).
\end{equation}

As the constant $C$ vanishes, 
we recover from (10) the usual Langevin equations (3) as a special case. 
In short, equations in (10), together with (11) and (12), are our 
generalized Langevin equations. 
\vspace{1pc}


\section{Our non-Markovian Smoluchowski equation}


Equations (10), (11) and (12) generate the following Fokker--Planck 
equation in phase space $(x,\bar p)$ (for details, see Appendix)
\begin{equation}
\frac{\partial {\cal F}}{\partial t}=-\frac{\partial (A_x{\cal
F})}{\partial x}- \frac{\partial (A_{\bar p}{\cal F})}{\partial
\bar p}+\frac{A_{xx}}{2} \frac{\partial^2 {\cal F}}{\partial
x^2}+ A_{x\bar p}\frac{\partial^2 {\cal F}}{\partial x\partial
\bar p}
 +\frac{A_{\bar p \bar p}}{2}\frac{\partial^2 {\cal F}}{\partial \bar p^2},
\end{equation}
where
$$
{\cal F}={\cal F}(x,\bar p,t)=\int D_{X\bar P\Psi}(x,\bar
p,\psi,t)d\psi.
$$
The quantities
$$
A_x=(\bar p/m)+ (C^2/t_c)^{1/3}
(e^{-t/t_c}-e^{-2t/t_c}),
$$
and
$$
A_{\bar p}=-\frac{dV}{dx} -(\gamma /m)\bar p+(CD/t_c)^{1/3}
(e^{-t/t_c}-e^{-2t/t_c})
$$
are the drift coefficients, whereas the time-dependent diffusion coefficients are given by
$$
A_{xx}=(C^2D)^{1/3}(1-e^{-t/t_c})^{2},
$$
$$
A_{x\bar p}=(D^2C)^{1/3}(1-e^{-t/t_c})^{2},
$$
and
$$
A_{\bar p \bar p}=D(1-e^{-t/t_c})^{2}.
$$
Combining $A_{xx}$, $A_{x\bar p}$, and $A_{\bar p \bar p}$ we notice that  
they satisfy the relation
\begin{equation}
\sqrt{A_{xx}A_{\bar p \bar p}}=A_{x\bar p}.
\end{equation}
Moreover, on replacing the 
Maxwell--Boltzmann (MB) distribution 
\begin{equation}
{\cal F}(x,\bar p)=\frac{1}{\sqrt{2\pi mk_B T}}
 e^{-(\bar {p}^2/2 m\kappa_B T)}e^{-(kx^2/2\kappa_B T)}
\end{equation}
into our Fokker--Planck equation (13) it is too easy to verify 
that (15) cannot become its solution. 
This means that our stochastic
process, described by (10--13), holds always away   from the thermal
equilibrium. That leads us to think that the physical meaning 
of the relation (14), which is a consequence of our 
assumption $\langle \Psi \rangle\neq 0$ in (11), is connected 
with nonequilibrium characteristics underlying the environment. 
In fact, as $C=0$ the constraint (14) is broken up and our generalized momentum 
$\bar P$ in Eq.(8) equals to $P$. Consequently, Eq.(13) reduces to the
non-Markovian Kramers equation in phase space $(x,p)$
\begin{equation}
\frac{\partial {\cal F}}{\partial t}=-\frac{p}{m}
\frac{\partial {\cal F}}{\partial x}+ \frac{\partial }{\partial p}
\left[\ \left(\ \frac{dV}{dx}+ \frac{\gamma}{m} p \right){\cal F}
\right]+ \frac{D(1-e^{-t/t_c})^{2}}{2}\frac{\partial^2 }{\partial
 p^2}{\cal F}.
\end{equation}
In the Markovian steady regime characterized by  
 $t\gg t_c$, or formally 
$t_c \rightarrow 0$, the MB distribution (15) with $\bar p=p$ 
turns out to be a solution to (16), thereby determining the
diffusion coefficient as being equal to $A_{pp}=D=2\gamma
\kappa_{B} T$. It is worth noting that according to
both the phase space equations (13) and (16) the following
fluctuating relation is valid
$$
\triangle X\triangle P>0,
$$
where $\triangle X=\sqrt{\langle X^2\rangle -\langle X\rangle^2}$,
and  $\triangle P=\sqrt{\langle P^2\rangle -\langle P\rangle^2}$. 
\vspace{1pc}

On the other hand, inserting
${\cal F}(x,\bar p,t)=f(x,t)\delta(\bar p)$ into (13) 
and taking into account the high friction condition
$$
\gamma \frac{\bar p}{m}=-\frac{dV}{dx},
$$
obtained from Newton's equations (5) on neglecting inertial effects 
($|d\bar p/dt|\ll|\gamma \bar p/m|$), we
arrive at  the non-Markovian Smoluchowski equation in position space
\begin{equation}
\frac{\partial f(x,t)}{\partial t}=-\frac{1}{\gamma}
\frac{\partial }{\partial x}[{\cal K}(x,t)f(x,t)]+
\frac{A_{xx}}{2}\frac{\partial^2 f(x,t)}{\partial x^2},
\end{equation}
where 
$$
{\cal K}(x,t)=-\frac{dV}{dx}+
\gamma(C^2/t_c)^{1/3}(e^{-t/t_c}-e^{-2t/t_c}).
$$
 Replacing (15) into (17) we obtain 
$A_{xx}=2\kappa_B T/\gamma$ in both
stationary and Markovian regimes.
\vspace{1pc}

Considering $V=0$ (free particle) and $C=0$ 
from our equation (12)  we derive the 
non-Markovian Rayleigh equation in  $p$-space 
\cite{Kampen81}
\begin{equation} 
\frac{\partial g(p,t)}{\partial t}=\frac{\gamma}{m}
\frac{\partial }{\partial p}[pg(p,t)]+
\frac{D(1-e^{-t/t_c})^2}{2}\frac{\partial^2 }{\partial p^2}g(p,t),
\end{equation}
with 
$$ 
g(p,t)=\int {\cal F}(x,p,t)dx.
$$
At thermal equilibrium  we find $D_{pp}=\gamma k_BT$
as being the diffusion coefficient in momentum space after inserting
(15) into (18). For both (16) and (18) the fluctuating relation
$$
\triangle X\triangle P=0
$$
is satisfied.
 \vspace{1pc}

From the mathematical viewpoint we note that    
we can derive the Kramers equation (16), the Smoluchowski equation (17), 
and the Rayleigh equation (18) as special cases of our equation of 
motion (13). That physically means that all the physics encapsulated 
into these equations of motion (16), (17), and (18) are in principle 
contained in our Eq.(13). 
 \vspace{1pc}

In the next sections we wish to survey the 
quantization of our Smoluchowki equation (17), thereby tackling the
problem of quantum tunneling at zero temperature in the strong
friction regime.


\section{Quantization of our Smoluchowski equation}


We start  with the non-Markovian Smoluchowski equation (17)
 at points $x_1$ and $x_2$; we multiply the first equation by
$f(x_2,t)$ and the second one by $f(x_1,t)$. We add the
resulting equations and obtain the following evolution equation
for the function $\xi=\xi(x_1,x_2,t)=f(x_1,t)f(x_2,t)$:
\begin{equation}
\frac{\partial \xi}{\partial t}=-\frac{1}{\gamma} \left(\
\frac{\partial}{\partial x_1} {\cal K}(x_1,t)+
\frac{\partial}{\partial x_2} {\cal
K}(x_2,t)\right)\xi+2A_{xx}\frac{\partial^2 \xi}{\partial p^2}.
\end{equation}
 We now perform the change of variables
 $$
 x=\frac{x_1+x_2}{2} \quad \quad, \quad \quad \eta=x_1-x_2.
 $$
We define our quantization process by introducing the following
Fourier transform \cite{Bolivar2004,Bolivar98,Bolivar2005}

\begin{equation}
F(x,p,t)=\frac{1}{2\pi \hbar}\int \xi(x,\eta,t)e^{\imath p \eta/\hbar}d\eta,
\end{equation}
$\hbar$ being the Planck constant $h$ divided by $2\pi$ and
$p=mdx/dt$ the  physical momentum. Inserting (20) into (19) we
arrive at our non-Gaussian, non-Markovian equation of motion in
quantum phase space $(x,p;\hbar)$
\begin{equation}
\frac{\partial F}{\partial t}=-\frac{1}{\gamma}({\cal O}F+{\cal A}F+{\cal B}F)+
\frac{{\cal D}(t)}{2}\frac{\partial^2 F}{\partial x^2}-2{\cal D}(t)\frac{p^2}{\hbar^2}F,
\end{equation}
with
\begin{equation}
{\cal
O}F=2\sum_{s=1,3,5,...}^{\infty}\frac{1}{(s-1)!(2\imath)^{s-1}}
\frac{\partial^s {\cal K}(x,t)}{\partial x^s}\frac{\partial^{s-1}
F}{\partial p^{s-1}},
\end{equation}
\begin{equation}
{\cal A}F=\sum_{r=0,2,4,...}^{\infty}\frac{1}{r!2^{r}}
\frac{\partial^r {\cal K}(x,t)}{\partial x^r}\frac{\partial^{r+1}
F}{\partial x \partial p^{r}},
\end{equation}
\begin{equation}
{\cal B}F=2\sum_{s=1,3,5,...}^{\infty}\frac{1}{s!2^s\imath^{s+1}}
\frac{\partial^s {\cal K}(x,t)}{\partial x^s}\left\{
s\frac{\partial^{s-1}F}{\partial p^{s-1}}+ p
\frac{\partial^{s}F}{\partial p^{s}} \right \},
\end{equation}
and
\begin{equation}
{\cal D}(t)=(DC^2)^{1/3}(1-e^{-t/t_c})^2.
\end{equation}
Our quantum Smoluchowski equation (21) is valid far from the
thermal equilibrium and for nonlinear external forces. We
emphasize that according to our quantization procedure, based on
the Fourier transformation (20), our Eq.(21) has to obey the
Heisenberg fluctuating relation
$$
\triangle X \triangle P\geq \frac{\hbar}{2},
$$
thus restoring the stochastic character of momentum variable in
the quantum domain. In the classical limit, $\hbar \rightarrow 0$,
we recover the classical expression $\triangle X \triangle P\geq
0$. Most specifically, assuming
$$
 F(x,p,t)=f(x,t)\delta(p)
$$
our quantum equation (21) leads to the classical Smoluchowski
equation (17), since
$$
\int {\cal O}Fdp=2\frac{\partial  {\cal K}}{\partial x}f(x,t),
$$
$$
\int {\cal A}Fdp={\cal K}\frac{\partial f(x,t) }{\partial x},
$$
$$
\int {\cal B}Fdp=-\frac{\partial  {\cal K}}{\partial x}f(x,t).
$$


\subsection{ Harmonic oscillator}


Let us consider $V=kx^2/2$, $k$ being a constant. Equation (21)
turns out to be written as
\begin{equation}
\frac{\partial F}{\partial t}=\frac{k}{\gamma}F+
\frac{k}{\gamma}[x+h(t)]\frac{\partial F}{\partial x}+\frac{{\cal D}(t)}{2}
\frac{\partial^2 F}{\partial x^2}-\frac{k}{\gamma}p\frac{\partial F}{\partial p}
-2{\cal D}(t)\frac{p^2}{\hbar^2}F,
\end{equation}
where
$$
h(t)= \frac{\gamma}{k}\left(\ \frac{C^2}{t_c}
\right)^{1/3}(e^{-2t/t_c}-e^{-t/t_c}).
$$
With the initial condition
\begin{equation}
F(x,p,t=0)=\frac{\sqrt{ab}}{\pi \hbar}e^{-ap^2/\hbar-bx^2/\hbar},
\end{equation}
$a$ and $b$ being Gaussian parameters, the solution for  (26)
reads
\begin{equation}
F(x,p,t)=\frac{1}{2\pi} \left(\ \frac{A(t)}{B(t)}\right)^{1/2}
 e^{-A(t)p^2-[x-c(t)]^2/4B(t)},
\end{equation}
where

\begin{equation}
A(t)= \left\{ \frac{a}{\hbar}-2\frac{g(t)}{\hbar^2} \right \}e^{-2kt/\gamma}+
\frac{2g(t)}{\hbar^2},
\end{equation}
\begin{equation}
B(t)= \left\{ \frac{\hbar}{4b}-\frac{g(t)}{2} \right \}e^{-2kt/\gamma}+
\frac{g(t)}{2},
\end{equation}
\begin{equation}
c(t)= ( 1-e^{-kt/\gamma}  )\mu(t),
\end{equation}
\begin{equation}
g(t)=\gamma (C^2 D)^{1/3} \left\{ \frac{1}{2k}-
2t_c\frac{e^{-t/t_c}}{2kt_c-\gamma}+
 \frac{t_ce^{-2t/t_c}}{2(kt_c-\gamma)}\right \},
\end{equation}
and
\begin{equation}
\mu(t)= \left(\ \frac{C}{t_c} \right)^{2/3} \left\{\frac{e^{-2t/t_c}}{2}-
 e^{-t/t_c}\right\}.
\end{equation}
Solution (28) leads to
\begin{equation}
\triangle X=\sqrt{2B(t)},
\end{equation}
\begin{equation}
\triangle P=\frac{1}{\sqrt{2A(t)}},
\end{equation}
that is,
\begin{equation}
\triangle X\triangle P=\sqrt{\frac{B(t)}{A(t)}}\geq
\frac{\hbar}{2}.
\end{equation}

\section{Quantum tunneling}

Now, let $t=\Delta \tau$ be a fixed time interval for 
observing the Brownian particle such that we
have a steady solution for (26), that is, $\partial F/\partial
t|_{t=\Delta\tau}=0$. During $\Delta \tau$ the system is therefore
stationary. In this context, we aim at to calculate the quantum
Kramers escape rate of a Brownian particle
 over a potential barrier in the
strong friction regime. \vspace{1pc}

We consider  a Brownian particle  moving in a double-well
potential $V(x)$. The barrier top is located  at point $x_b$,
while the two bottom wells is at $x_a$ and $x_c$, such that
$V(x_a)=V(x_c)=0$, $x_a<x_b$. The starting point is our  solution
(28) we modify according to
\begin{equation}
F(x,p,\Delta \tau)=\alpha \phi(x,p)
 e^{-A(\Delta \tau)p^2-[x-c(\Delta \tau)]^2/4B(\Delta \tau)},\quad
 \alpha=constant.
\end{equation}
Inserting (37) into the steady version of (26) we derive the
ordinary differential  equation for the function $\phi(x,p)$
\begin{equation}
 \frac{{\cal D}\gamma}{2k}\frac{d^2\phi}{d\xi^2}=
\xi \frac{d\phi}{d\xi}, \quad \quad \xi=p-x,
\end{equation}
since
\begin{equation}
B(\Delta \tau)=\frac{{\cal D}\gamma}{4k},
\end{equation}
\begin{equation}
A(\Delta \tau)=\frac{{\cal D}\gamma}{k\hbar^2},
\end{equation}
\begin{equation}
h(\Delta \tau)=-c(\Delta \tau).
\end{equation}
From (39) and (40) we determine the parameters of Gaussian
function (27) as being
\begin{equation}
a=\frac{{\cal D}\gamma}{\hbar}\left\{
\frac{1}{k}+2u\left(1-e^{k\Delta \tau/\gamma} \right) \right\},
\end{equation}
and
\begin{equation}
b=\frac{\hbar}{{\cal D}\gamma}\frac{1}{\left\{
\frac{1}{k}+2u\left(1-e^{k\Delta \tau/\gamma} \right) \right\}},
\end{equation}
with
\begin{equation}
u=u(\Delta \tau)= \left\{- 2t_c\frac{e^{-\Delta
\tau/t_c}}{2kt_c-\gamma}+
 \frac{t_ce^{-2\Delta \tau/t_c}}{2(kt_c-\gamma)}\right \}.
\end{equation}
It follows then that $ab=1$, whereas from the identity (41) we
have the following relation among the time scales $\Delta
\tau$ (the observation time), $t_c$ (the correlation time),
and $t_r=\gamma/k$ (the relaxation time):
\begin{equation}
t_c^{1/3}=\frac{(e^{- \Delta \tau/t_r}-1)}{2t_r}
 \frac{(e^{-\Delta \tau/t_c}-2)}{(e^{-\Delta \tau/t_c}-1)}.
\end{equation}
\vspace{1pc}

A solution to the differential equation (38) is given by
\begin{equation}
\phi(\xi)=\sqrt{\frac{-k}{\pi{\cal D}\gamma}}
\int_{-\infty}^{\xi}e^{(k/{\cal D}\gamma)\xi^2} d\xi, \quad \quad
k<0,
\end{equation}
wherein we have used the boundary condition $\phi(\xi\rightarrow
+\infty)=1$. This result
 confirms the fact that the region around the barrier at $x_b$, in which the
curvature of the potential is negative, is quite relevant in the
calculation of the diffusion current, as we will see below.
Substituting (46), (39), and (40) into (37) and expanding the
ensuing $F(x,p)$ around $x_b$ we obtain
\begin{equation}
F=\alpha e^{-2V(x_b)/{\cal D}\gamma} \sqrt{\frac{m\omega^{2}_{b}}{
\pi{\cal D}\gamma}}e^{-({\cal D}\gamma/m\omega_{b}^{2}\hbar^2)p^2
+(m\omega_{b}^{2}/{\cal
D}\gamma)(x-x_b-c)^2}\\
\int_{-\infty}^{\xi}e^{-(m\omega_{b}^{2}/{\cal D}\gamma)\xi^2}
d\xi,
\end{equation}
with $k_b=-m\omega^{2}_{b}$. At point $x=x_b$ we find the
following diffusion current
\begin{equation}
j_b=\int F(x=x_b,p)\frac{p}{m}dp= \frac{\alpha
\omega^{2}_{b}\hbar^2 e^{(m\omega_b^2c^2/ {\cal D}\gamma)}}
{2{\cal D}\gamma\sqrt{1-({\cal D}\gamma/m\omega^{2}_{b}\hbar)^2 }}
e^{-2V(x_b)/{\cal D}\gamma},
\end{equation}
where
\begin{equation}
c=c(\Delta \tau)= ( 1-e^{m\omega^2_b\Delta \tau/\gamma}  )\left(\
\frac{C}{t_c} \right)^{2/3} \left\{\frac{e^{-2\Delta
\tau/t_c}}{2}-
 e^{-\Delta \tau/t_c}\right\}.
\end{equation}
Our result (48) is valid provided
\begin{equation}
m\omega^{2}_{b}\hbar>{\cal D}\gamma.
\end{equation}
\vspace{1pc}

At the vicinity of  $x_a$ we cannot use the stationary solution
(47), since it is only valid for negative curvature $k<0$, hence
we use the (nonnormalized) function
$$
F(x,p)=\alpha
 e^{-({\cal D}\gamma/k\hbar^2)p^2-(k/{\cal D}\gamma)(x-c)^2}
$$
 to finding  the number of
Brownian particles injected around $x_a$:
$$
\nu_a=\int \int F(x,p)dxdp=\alpha \pi\hbar.
$$
Using $\nu_a$ and the current (48)
the non-Markovian quantum Kramers escape rate at non-equilibrium
regime reads
\begin{equation}
\Gamma=\frac{j_a}{\nu_a}= \frac{\omega^{2}_{b}\hbar
e^{m\omega_b^2c^2/ {\cal D}\gamma}} {2\pi{\cal
D}\gamma\sqrt{1-({\cal D}\gamma/m\omega^{2}_b\hbar)^2 }}
e^{-2V(x_b)/{\cal D}\gamma}.
\end{equation}
 In the Markovian limit $\Delta \tau \gg t_c$, or formally
$t_c\rightarrow 0$, from (51) we obtain
\begin{equation}
\Gamma= \frac{\omega^{2}_{b}\hbar} {2\pi{\cal
D}\gamma\sqrt{1-({\cal D}\gamma/m\omega^{2}_b\hbar)^2 }}
e^{-2V(x_b)/{\cal D}\gamma}.
\end{equation}
Non-Markovian properties are therefore responsible for the 
enhancement of the quantum tunneling rate far from the thermal equilibrium.

\vspace{1pc}

We want now to evaluate the diffusion coefficient ${\cal D}$
present in (51) or (52) using the well-established principles of
equilibrium thermodynamics. To this end, let us assume that during
$\Delta \tau$ our open system has attained a thermal equilibrium
situation in which is valid  the principle of energy equipartition
\begin{equation}
\langle E\rangle= \frac{\langle P^2
\rangle}{2m}+\frac{k}{2}\langle X^2\rangle=\kappa_B T
\end{equation}
 that associates the  stochastic dynamics of the Brownian particle
 (e.g., its average total energy) to
 equilibrium thermodynamics underlying the thermal environment (Boltzmann's
 constant $\kappa_B$ and temperature $T$).
\vspace{1pc}

Replacing (39) and (40) into (34) and (35), respectively, we find
\begin{equation}
\langle X^2\rangle= \frac{ {\cal D}\gamma }{ 2k },
\end{equation}
\begin{equation}
\langle P^2\rangle=\frac{  k\hbar^2  }{  2\gamma {\cal D}  }
\end{equation}
that lead to the following equation according to (53)
\begin{equation}
 {\cal D}^2-\frac{4\kappa_B T}{\gamma}{{\cal
 D}}+\frac{k\hbar^2}{m\gamma^2}=0,
\end{equation}
 whose solution is the quantum diffusion coefficient \cite{Comment}
\begin{equation}
{\cal D}=\frac{2\kappa_B T}{\gamma} +\frac{1}{\gamma}
\sqrt{(2\kappa_BT)^2-(\hbar\Omega)^2}; \quad \Omega^2=k/m, \quad
k>0,
\end{equation}
valid for any finite temperature $T\geq \hbar \Omega/2\kappa_B$,
or
\begin{equation}
{\cal D}=\frac{2\kappa_B T}{\gamma} +\frac{1}{\gamma}
\sqrt{(2\kappa_BT)^2+(\hbar\omega)^2}; \quad \omega^2=-k/m, \quad
k<0,
\end{equation}
valid for any finite temperature $T\geq 0$. At high temperature,
$\kappa_BT\gg\hbar\Omega,\hbar\omega$, both (57) and (58) lead to
the classical diffusion coefficient ${\cal D}=4\kappa_B T/\gamma$.
On the other hand, at low temperature so that the quantum energy
of the Brownian particle is equal to the thermal energy of the
reservoir, i.e., $\kappa_BT=\hbar\Omega/2=\hbar\omega/2$, (57) and
 (58) lead to ${\cal D}=\hbar \Omega/\gamma$
and ${\cal D}=\hbar \omega(1+\sqrt2)/\gamma$, respectively. At
lower temperature,  $\kappa_B T\ll \hbar \omega$, from (58) we
obtain ${\cal D}=(\hbar \omega+2\kappa_BT)/\gamma$ which in turn
leads to ${\cal D}=\hbar \omega/\gamma$ at zero
temperature.\vspace{1pc}

Due to the condition (50) the non-Markovian quantum Kramers escape
rate is thereby constrained to the low-temperature realm
($T<\hbar\omega_b/\kappa_B$)
\begin{equation}
\Gamma=\frac{\omega^{2}_{b}\hbar e^{m\omega_b^2c^2/
(\omega_b\hbar+2\kappa_BT)} }{2\pi
(\omega_b\hbar+2\kappa_BT)\sqrt{1-[(\omega_b\hbar+2\kappa_BT)/m\omega^{2}_{b}\hbar]^2
} } e^{-2V(x_b)/(\omega_b\hbar+2\kappa_BT)}
\end{equation}
that in the Markovian limit  leads to
\begin{equation}
\Gamma=\frac{\omega^{2}_{b}\hbar }{2\pi
(\omega_b\hbar+2\kappa_BT)\sqrt{1-[(\omega_b\hbar+2k_BT)/m\omega^{2}_{b}\hbar]^2
} } e^{-2V(x_b)/(\omega_b\hbar+2\kappa_BT)}.
\end{equation}
At zero temperature from (59) we obtain the non-Markovian rate

\begin{equation}
\Gamma=\frac{\omega_{b}e^{m\omega_bc^2/\hbar}} {2\pi
\sqrt{1-(1/m\omega_b)^2}  } e^{-2V(x_b)/\omega_b\hbar},
\end{equation}
whereas in the Markovian case we derive
\begin{equation}
\Gamma=\frac{\omega_{b}} {2\pi \sqrt{1-(1/m\omega_b)^2} }
e^{-2V(x_b)/\omega_b\hbar}
\end{equation}
that is the same result obtained in our previous work
\cite{Bolivar2005}. \vspace{1pc}

We wish to point out that our  Markovian results (60) and (62) are
independent of the friction constant $\gamma$ at thermal
equilibrium. This means that over the barrier at low temperatures
(including zero temperature) Brownian particles may decay into a
metastable well around $x_c$ at high temperatures. This may take
place without violating the Second Law of Thermodynamics since in
the quantum domain at low temperatures the particles could flux
overcoming any dissipative mechanism (superfluidity phenomenon!).
\vspace{1pc}

So far we have considered the thermal environment as having a
general nature. Supposing then that the thermal reservoir consists
of a set of many harmonic oscillators we can generalize the
equipartition theorem of energy (53) taking into account the
quantum nature of the heat bath:
\begin{equation}
\langle E\rangle= \frac{\langle P^2
\rangle}{2m}+\frac{k}{2}\langle X^2\rangle=(\hbar\nu/2)
\coth(\hbar \nu/2\kappa_B T).
\end{equation}
[$\nu=(k/m)^{1/2}$ for $k>0$, and $\nu=(-k/m)^{1/2}$ for $k<0$].
 It follows then the equation
 ${\cal D}^2-4\langle E\rangle{\cal D}/\gamma+k\hbar^2/m\gamma^2=0$
 that yields the solution
\begin{equation}
{\cal D}=\frac{\hbar\Omega}{\gamma}\coth\frac{\hbar
\Omega}{2\kappa_BT}
 +\frac{\Omega \hbar}{\gamma}
\sqrt{\coth^2\frac{\hbar \Omega}{2\kappa_BT}-1}; \quad
\Omega^2=k/m, \quad k>0,
\end{equation}
 or
\begin{equation}
{\cal D}=\frac{\hbar\omega}{\gamma}\coth\frac{\hbar
\omega}{2\kappa_BT}
 +\frac{\omega \hbar}{\gamma}
\sqrt{\coth^2\frac{\hbar \omega}{2\kappa_BT}+1}; \quad
\omega^2=-k/m, \quad k<0.
\end{equation}
 At high temperature,
$\kappa_BT\gg\hbar\Omega,\hbar\omega$, both (64) and (65) lead to
${\cal D}=4\kappa_B T/\gamma$. At low temperature,
$\kappa_BT\ll\hbar\Omega, \hbar \omega$, (including zero
temperature $T=0$), (64) and (65) lead to ${\cal D}=\hbar
\Omega/\gamma$ and ${\cal D}=\hbar \omega(1+\sqrt2)/\gamma$,
respectively. Hence we obtain the non-Markovian quantum Kramers
escape rate (52) as being given by the expression
\begin{equation}
\Gamma= \frac{\omega_{b} e^{   m\omega_b^2c^2/ \hbar
\omega_b(1+\sqrt2)}}
{2\pi(1+\sqrt{2})\sqrt{1-[(1+\sqrt2)/m\omega_b]^2 }}
e^{-2V(x_b)/\hbar \omega_b(1+\sqrt2)}.
\end{equation}
In the Markovian limit we obtain
\begin{equation}
\Gamma= \frac{\omega_{b}}
{2\pi(1+\sqrt{2})\sqrt{1-[(1+\sqrt2)/m\omega_b]^2 }}
e^{-2V(x_b)/\hbar \omega_b(1+\sqrt2)}
\end{equation}
 that exhibits the influence of the zero-point energy of the
thermal reservoir upon the quantum tunneling rate of our Brownian
particle at zero temperature. Eq.(67) is also independent of
frictional constant $\gamma$, thus leading to the superfluidity
phenomenon.

\section{Summary and discussions}

In this paper we have addressed the issue of quantum tunneling at
low (including zero temperature) for non-Markovian open systems
away from the thermal equilibrium. From our quantum Smoluchowski
equation (21) we have derived the non-equilibrium, non-Markovian
quantum escape rate (51) that does depend on  both the friction
and diffusion coefficients.\vspace{1pc}

We have thus provided an alternative method of quantizing the
Smoluchowski equation taking into account non-Markovian and
non-equilibrium effects. This result, coming from our
non-Hamiltonian account, is novel and is not present in others
approaches
\cite{Ankerhold2005,Banerjee2003,Ankerhold2005a,Vacchini2002}.
\vspace{1pc}

At low temperature and in the Markovian regime our quantum Kramers
escape rates (60) and (67) are independent of damping constant
$\gamma$. That is, dissipation has no effect on the quantum
tunneling, thus giving rise to superfluidity. This result has been
misunderstood by Ankerhold et al.\cite{Ankerhold2005a} that state:
``This finding (...) contracts existing theoretical results
verified by experimental data''. This statement is incorrect since
Ao et al.\cite{Ao94} have arrived at our same conclusion by
studying Landau--Zener tunneling in a dissipative environment:
``no effect of dissipation is present at zero temperature''.
\vspace{1pc}

We hope our present work could foster development of
experimental researches on quantum tunneling in overdamped open
systems in order that we can compare our theoretical predictions
with experimental data.\vspace{1pc}

In a forthcoming paper \cite{Bolivar2006} we pursue our
non-Hamiltonian approach in surveying quantum tunneling effects
from the quantization of our non-equilibrium, non-Markovian
Fokker--Planck equation (13) and of the Rayleigh equation (18), as
well as quantum tunneling in the small friction regime (energy
diffusion). 
\vspace{1pc}

To conclude, we want to point out that non-linear effects due to 
the external potential can
be also analyzed through our quantum Smoluchowski equation (21)
\begin{equation}
\frac{\partial F}{\partial t}= \frac{1}{\gamma}\left(\
x^2\frac{\partial F}{\partial x}+ \frac{1}{4}\frac{\partial^3
F}{\partial x\partial p^2}+2 xF- 2 xp\frac{\partial F}{\partial p}
\right)+ \frac{{\cal D}(t)}{2}\frac{\partial^2 F}{\partial x^2}-
2{\cal D}(t)\frac{p^2}{\hbar^2}F
\end{equation}
for the cubic potential $V=x^3/3$, for instance. \vspace{20pc}

\section*{Acknowledgements}

I  wish to thank Dr. Ping Ao (University of Washington) for
bringing to my attention his papers present in Ref.\cite{Ao94} and 
Dr. Annibal Figueiredo for the scientific support.
\vspace{35pc}


\section*{Appendix: Derivation of our Fokker--Planck equation [Eq.(13)]}


In this appendix  we want to show in somewhat details how we could explicitly construct 
the Fokker--Planck equation (13) from the system of stochastic differential 
equations \cite{Bolivar2004} 
\begin{equation}
\frac{d\bar P}{dt}= -\frac{dV}{dX}-\frac{\gamma}{m}\bar P+b_1\Psi
\quad;\quad \frac{dX}{dt}=\frac{\bar P}{m}+b_2\Psi.
\end{equation}
Equations (69) yield the results
\begin{equation}
 \Delta \bar P =- \left(\ \frac{dV}{dX}+\frac{\gamma}{m}\bar P\right)\Delta t+
\int_{t}^{t+\Delta t}
b_1(t')\langle \Psi(t')\rangle dt'
\end{equation}
and
\begin{eqnarray}
 \Delta  X =\frac{\bar P}{m}\Delta t- \left(\ \frac{dV}{dX}+
 \frac{\gamma}{m}\bar P\right)\frac{(\Delta t)^2}{m}+\frac{1}{m}
\int_{t}^{t+\Delta t}\int _{t}^{s}b_1(t')\langle \Psi(t')
\rangle dt'ds+
\nonumber \\
\int_{t}^{t+\Delta t}b_2(t')\langle \Psi(t')\rangle dt'.
\end{eqnarray}
Using $\Delta \bar P=\bar P(t+\Delta t)-\bar P(t)$ and 
$\Delta X=X(t+\Delta t)-X(t)$ we calculate the following quantities

\begin{equation}
A_{\bar p}=\lim_{ \Delta t \rightarrow 0}\frac{\langle \Delta \bar P \rangle}
{\Delta t}=-\frac{dV}{dX}-\frac{\gamma}{m}\bar P+
\lim_{ \Delta t \rightarrow 0}\frac{1}{\Delta t}\int_{t}^{t+\Delta t}
b_1(t')\langle \Phi(t')\rangle dt',
\end{equation}

\begin{eqnarray}
A_{p\bar p}=\lim_{ \Delta t \rightarrow 0}\frac{\langle (\Delta \bar P)^2 \rangle}
{\Delta t}=-2\left(\ \frac{dV}{dX}+\frac{\gamma}{m}\bar P\right)
\lim_{ \Delta t \rightarrow 0}\int_{t}^{t+\Delta t}
b_1(t')\langle \Psi(t')\rangle dt'+
\nonumber \\
\lim_{ \Delta t \rightarrow 0}\frac{1}{\Delta t}
\int_{t}^{t+\Delta t}\int_{t}^{t+\Delta t}
b_1(t')b_1(t'')\langle \Psi(t')\Psi(t'')\rangle dt'dt'',
\end{eqnarray}
\begin{eqnarray}
A_{x}=\lim_{ \Delta t \rightarrow 0}\frac{\langle \Delta X \rangle}
{\Delta t}=\frac{\bar P}{m}+\frac{1}{m}
\lim_{ \Delta t \rightarrow 0}\int_{t}^{t+\Delta t}\int_{t}^{s}
b_1(t')\langle \Psi(t')\rangle dt'ds+
\nonumber \\
\lim_{ \Delta t \rightarrow 0}
\int_{t}^{t+\Delta t}b_2(t')\langle \Psi(t')\rangle dt',
\end{eqnarray}
\begin{equation}
A_{xx}=\lim_{ \Delta t \rightarrow 0}\frac{\langle (\Delta X)^2 \rangle}
{\Delta t}=I_1+I_2+I_3+I_4
\end{equation}
with

\begin{equation}
I_{1}=\frac{2\bar P}{m^2}\int_{t}^{t+\Delta t}\int_{t}^{s}
b_1(t')\langle \Psi(t')\rangle dt'ds,
\end{equation}
\begin{equation}
I_{2}=\frac{2\bar P}{m}\int_{t}^{t+\Delta t}b_2(t')\langle \Psi(t')\rangle dt',
\end{equation}
\begin{equation}
I_{3}=\frac{1}{m^2}\lim_{ \Delta t \rightarrow 0}
\int_{t}^{t+\Delta t}\int_{t}^{t+\Delta t}\int_{t}^{r}\int_{t}^{s}
b_1(t')b_1(t'')\langle \Psi(t')\Psi(t'')\rangle dt'dt''drds,
\end{equation}

\begin{equation}
I_{4}=\frac{2}{m}\lim_{ \Delta t \rightarrow 0}
\int_{t}^{t+\Delta t}\int_{t}^{t+\Delta t}\int_{t}^{s}
b_1(t')b_2(t'')\langle \Psi(t')\Psi(t'')\rangle dt'dt''ds,
\end{equation}

\begin{equation}
I_{5}=\int_{t}^{t+\Delta t}\int_{t}^{t+\Delta t}
b_1(t')b_2(t'')\langle \Psi(t')\Psi(t'')\rangle dt'dt'',
\end{equation}

and 
\begin{equation}
A_{x\bar p}=\lim_{ \Delta t \rightarrow 0}\frac{\langle \Delta X \Delta \bar P\rangle}
{\Delta t}=\xi_1+\xi_2+\xi_3+\xi_4+\xi_5,
\end{equation}
where 
\begin{equation}
\xi_1=\frac{\bar P}{m}\int_{t}^{t+\Delta t}
b_1(t')\langle \Psi(t')\rangle dt',
\end{equation}
\begin{equation}
\xi_2=-\frac{1}{m}\left(\ \frac{dV}{dX}+\frac{\gamma}{m}\bar P\right)
\int_{t}^{t+\Delta t}\int_{t}^{s}
b_1(t')\langle \Psi(t')\rangle dt'ds,
\end{equation}
\begin{equation}
\xi_3=\frac{1}{m}\lim_{ \Delta t \rightarrow 0}\frac{1}{\Delta t}
\int_{t}^{t+\Delta t}\int_{t}^{t+\Delta t}\int_t^s
b_1(t')b_1(t'')\langle \Psi(t')\Psi(t'')\rangle dt'dt''ds,
\end{equation}
\begin{equation}
\xi_4=-\frac{\bar P}{m}\int_{t}^{t+\Delta t}
b_2(t')\langle \Psi(t')\rangle dt'
\end{equation}
\begin{equation}
\xi_5=\lim_{ \Delta t \rightarrow 0}\frac{1}{\Delta t}
\int_{t}^{t+\Delta t}\int_{t}^{t+\Delta t}
b_2(t')b_2(t'')\langle \Psi(t')\Psi(t'')\rangle dt'dt''.
\end{equation}
\vspace{1pc}

After using our definitions 

\begin{equation}
\langle \Psi(t')\Psi(t'')\rangle =(D/t^2_c)^{1/3}e^{-(t''-t')/t_c};
\quad \quad \langle \Psi \rangle=(C/t_c^2)^{1/3}e^{-t'/t_c},
\end{equation}
and
\begin{equation}
b_1=\int_{0}^{t}
\langle \Psi(t')\Psi(t'')\rangle dt''=(Dt_c)^{1/3}(1-e^{-t/t_c}),
\end{equation}
\begin{equation}
b_2=\int_{0}^{t}
\langle \Psi(t')\rangle dt'=(Ct_c)^{1/3}(1-e^{-t/t_c}).
\end{equation}
into (72--75) and (81) we obtain our Fokker--Planck equation (13) with 
the coefficients
\begin{equation}
A_x= (\bar p/m)+ \left(\ \frac{C^2}{t_c} \right)^{1/3}
(e^{-t/t_c}-e^{-2t/t_c}),
\end{equation}
\begin{equation}
A_{\bar p}=-kx -\left(\ \frac{\gamma}{m} \right)\bar p+\left(\ \frac{CD}{t_c} \right)^{1/3}
(e^{-t/t_c}-e^{-2t/t_c}),
\end{equation}
\begin{equation}
A_{xx}=(C^2D)^{1/3}(1-e^{-t/t_c})^{2},
\end{equation}
\begin{equation}
A_{x\bar p}=(D^2C)^{1/3}(1-e^{-t/t_c})^{2},
\end{equation}
\begin{equation}
A_{\bar p \bar p}=D(1-e^{-t/t_c})^{2}.
\end{equation}
\vspace{28pc}




\begin{thebibliography}{99}
\bibitem{Comment}
Solution ${\cal D}=2\kappa_BT/\gamma-(1/\gamma)
\sqrt{(2\kappa_BT)^2-(4k\hbar^2/m)}$ is
not considered because it predicts no diffusion (${\cal D}=0$) in the classical
limit $\hbar \rightarrow 0$.
\end{thebibliography}
\end{document}